\documentclass{article}
\usepackage{epsfig}
\newcommand{\bfr}{\begin{flushright}}
\newcommand{\efr}{\end{flushright}}
 
\begin{document}
\title{Quantum Effects near Charged Dilatonic
Black Holes
}
\author{Kiyoshi Shiraishi\\
Akita Junior College, Shimokitade-Sakura, Akita-shi, \\Akita 010,
Japan
}
\date{Modern Physics Letters {\bf A7}, No. 38 (1992) pp. 3569--3574
}
\maketitle
\begin{abstract}
The expectation value $\langle\varphi^2\rangle$ in the Hartle-Hawking
vacuum is calculated for minimally-coupled neutral scalar fields at the
horizon of a charged dilatonic black hole.
\end{abstract}

\bigskip

Some exotic classes of black holes have attracted much attention of
modern theoretical physicists. For example, black holes with quantum
hair are discovered recently.\cite{1,2,3,4,5} The thermodynamical
properties of the black holes are expected to receive some corrections
at the presence of quantum hair. As another example, we mention the
study on the properties of dilatonic black holes in two-dimensional
space-time.\cite{6} The black holes in two dimensions are known to be
concerned with conformal field theory.\cite{6} Furthermore, they are
useful tools for the nonperturbative study on the evaporation of black
holes.\cite{7,8}

A charged, dilatonic black hole in four (or more) dimensions is now a
very popular object \cite{9,10}; various aspects of their classical and
quantum properties are curious and worth
studying.\cite{11,12,13,14,15,16,17,18,19} In this paper, we treat this
charged dilatonic black hole in four dimensions. The first motivation to
introduce the dilatonic black holes originates from string theory. The
interpolation between the stringy (dilatonic) electromagnetism and the
usual Maxwell system has been considered by introducing an arbitrary
value for dilaton coupling ``$a$'' to the Maxwell field. The action
which governs the system can be written as
\begin{equation}
S=\int d^4x\frac{\sqrt{-g}}{16\pi}[R-2(\nabla\phi)^2-e^{-2a\phi}F^2]\,,
\end{equation}
where $R$ is the scalar curvature, $\phi$ is the dilaton and $F$ is the
field strength of the Maxwell field. The special choice $a =1$ does
correspond to the string effective action. The study on the system with
arbitrary values for dilaton coupling shows clearly the effect of
adding the dilaton.

In this paper, we study the vacuum polarization in a background of the
charged dilatonic black holes. This is expected to be closely connected
to the thermal description of the black holes. We compute the quantity
$\langle\varphi^2\rangle$ for minimally-coupled real scalar fields at
the horizon, as the simplest example.

The metric for a spherically symmetric charged dilatonic black hole in
the Einstein-Maxwell-dilaton system with an arbitrary dilaton coupling,
$a$, is written in the form \cite{9,10}
\begin{equation}
ds^2=-\Delta\sigma^{-2}dt^2+\sigma^2(\Delta^{-1}dr^2+r^2d\Omega^2_2)\,,
\label{eq2}
\end{equation}
where
\begin{equation}
\Delta(r)=\left(1-\frac{r_+}{r}\right)\left(1-\frac{r_-}{r}\right)
\quad \mbox{and}\quad
\sigma^2(r)=\left(1-\frac{r_-}{r}\right)^{2a^2/(1+a^2)}\,.
\label{eq3}
\end{equation}
Here $d\Omega^2_2$ is the standard line element for a sphere with unit
radius. 

In these expressions $r_+$  and $r_-$ are integration constants
and they are connected to the mass $M$ and electric charge $Q$ of the
black hole through the relations
\begin{equation}
2M=r_++\frac{1-a^2}{1+a^2}r_-\quad\mbox{and}\quad
Q^2=\frac{r_+r_-}{1+a^2}\,,
\label{eq4}
\end{equation}
the horizon length is $r_+$.

The configurations of the dilaton $\phi$ and electric fields are given
by
\begin{equation}
e^{2a\phi}=\sigma^2(r)\quad\mbox{and}\quad F=\frac{Q}{r^2}dt\wedge dr\,,
\end{equation}
respectively.

One can easily see that the geometry of the fields reduces to the Reissner-
Nordstr\"om space-time when $a=0$. The case with $a=1$ corresponds to
string theory.

Now let $\varphi$ be a real scalar field which couples {\it minimally}
to the scalar curvature of the space-time. Here we treat the
Hartle-Hawking vacuum.\cite{20} We take the vacuum polarization
$\langle\varphi^2\rangle$ as the coincident limit of the two-point
function $\langle\varphi(x)\varphi(x')\rangle$ with proper
regularization.\cite{21,22,23,24,25,26,27,28,29,30} Then the two-point
function of the scalar field can be obtained most clearly by the
mode-sum method. We have to find the solution for the wave equation in
a background geometry of the charged dilatonic black hole (\ref{eq2})
with Euclidean signature.

If it is required that there is no singularity at the horizon $r=r_+$
in the Euclideanized geometry, the Euclidean time $\tau$ must be
periodic with period $1/T_H$, where $T_H$ is the Hawking temperature
\cite{9,10,11,12}
\begin{equation}
T_H=\frac{1}{4\pi
r_+}\left(1-\frac{r_-}{r_+}\right)^{(1-a^2)/(1+a^2)}\,.
\end{equation}

A mode function for the solution of the massless Klein-Gordon equation can be
written as
\begin{equation}
\varphi_{nlm}=R_{nl}(r)Y_{lm}(\theta,\varphi)e^{-i2\pi nT_H \tau} \,,
\end{equation}
where $Y_{lm}(\theta,\varphi)$ is the spherical harmonics.

To make the differential equation for the radial function a simpler
form, we use a new variable $y$ defined by the relation
\begin{equation}
y=r-r_-\,.
\end{equation}

Then the radial function is the solution of the following differential
equation:
\begin{eqnarray}
&
&\frac{d}{dy}\left\{y^2\left(1-\frac{y_H}{y}\right)\frac{dR(y)}{dy}\right\}\nonumber
\\ & &\qquad-(2\pi
nT_H)^2y^2\frac{\left(1+\frac{r_-}{y}\right)^{2/(1+a^2)}}{1-\frac{y_H}{y}}R(y)-l(l+1)R(y)=0\,,
\label{eq9}
\end{eqnarray}
where $y_H=r_+-r_-$.

One can immediately realize that the solution for (\ref{eq9}) will have
non-analytic expression in general. Only $n=0$ mode is exactly soluble
in the form of the special function.

The $n=0$ mode takes exactly the same form as in the case with the
Schwarzchild space time treated in Ref.~\cite{26} by Frolov et al. Of
course, the higher modes are quite different. But, if we need only the
two-point Hamilton one of which point lies on the horizon membrane
($r=r_+$), the contribution from the higher modes becomes irrelevant
for the situation.

Thus the procedure from now on must be similar to the method adopted by
Frolov et al.\cite{26} The quantity
$\langle\varphi(x)\varphi(x')\rangle$ behaves like that in Frolov's
case as long as the point $x$ lies on the horizon membrane. The only
difference is the overall coefficient proportional to the Hawking
temperature $T_H$. Therefore, it depends on the value of dilaton
coupling only through the coefficient. When the separation is
restricted to the radial direction, the two-point function takes the
form,
\begin{equation}
\langle\varphi(x)\varphi(x')\rangle=\frac{\left(1-\frac{r_-}{r_+}
\right)^{-2a^2/(1+a^2)}}{16\pi^2r_+^2\sinh^2\eta}
\end{equation}
where $x=(\tau, y_H,\theta,\varphi)$ and $x'=(\tau,
y,\theta,\varphi)$. $\eta$ is defined by
\begin{equation}
\cosh \eta = \left(\frac{y}{y_H}\right)^{1/2}\,.
\end{equation}

To compute the vacuum polarization $\langle\varphi^2\rangle(x)$, where
$x$ is located on the horizon, we must subtract the divergence part of
the two-point function. The higher order in the Schwinger-de Witt
expansion of the propagator (two-point function) with respect to the
powers of the geodesic distance between the two points are not yet
known; moreover, as the present case, the coefficients for the
expansion takes awkward forms for non-conformally-coupled fields.
Fortunately, the scalar curvature vanishes at the horizon $r=r_+$ ($y=
y_H; \eta=0$) of the charged dilatonic black hole in four dimensions.
(This does not hold in other dimensions.)

Taking such things into consideration, the divergent and constant contributions
in the Schwinger-de Witt expansion near the horizon is written by the use of the
geodesic distance and the Ricci curvature as \cite{25}
\begin{equation}
\langle\varphi(x)\varphi(x')\rangle_{div}=\frac{\Delta^{1/2}}{8\pi^2
\sigma(x,x')}=\frac{1}{8\pi^2
\sigma(x,x')}\left(1+\frac{1}{12}R_{\alpha\beta}\sigma^{,\alpha}
\sigma^{,\beta}\right)\,,
\end{equation}
where $\sigma=s^2/2$ and $s(x,x')$ is the geodesic distance between
$x$ and $x'$. For radial separation, $s$ is given by
\begin{eqnarray}
&&s=\int_{y_H}^y\frac{\left(1+\frac{r_+}{y}\right)^{1/(1+a^2)}}
{\sqrt{1-\frac{y_H}{y}}}\nonumber \\
&&=2y_H\left(1-\frac{r_-}{r_+}\right)^{-1/(1+a^2)}\int_0^\eta               
\cosh^2\eta\left(1-\frac{r_-}{r_+}\tanh^2\eta\right)^{1/(1+a^2)}d\eta\nonumber
\\ &&=2r_+\left(1-\frac{r_-}{r_+}\right)^{-a^2/(1+a^2)}
\left(\eta+\frac{1}{3}\left(1-\frac{r_-}{(1+a^2)r_+}\right)
\eta^3+O(\eta^5)\right)\,,
\end{eqnarray}
and
\begin{equation}
R_{\alpha\beta}\frac{\sigma^{,\alpha}\sigma^{,\beta}}{\sigma}=
2R^y_y=-\frac{2Q^2}{r_+^4}\left(1-\frac{r_-}{r_+}\right)^{-2a^2/(1+a^2)}\,.
\end{equation}

Using these and (\ref{eq4}), we get the final result:
\begin{eqnarray}
\langle\varphi^2\rangle_{horizon}&=&\lim_{x'\rightarrow x}
(\langle\varphi(x)\varphi(x')\rangle-
\langle\varphi(x)\varphi(x')\rangle_{div})\nonumber
\\
&=&\frac{1}{48\pi^2r_+^2}\left(1-\frac{r_-}{(1+a^2) r_+}\right)
\left(1-\frac{r_-}{r_+}\right)^{-2a^2/(1+a^2)}\,.
\end{eqnarray}

We display the result in Figs. 1 and 2 in two different manners. 
Figures $1a,b$
show $16\pi^2M^2\langle\varphi^2\rangle$ at the horizon versus
$q^2=Q^2/\{(1+a^2)M^2\}$ for
$a^2=0, 1/3, 1, 3$, while Fig. 2 shows
$\langle\varphi^2\rangle/(4T_H^2)$ at the horizon versus $q^2$ for $a^2
=0, 1/3, 1, 3$.

\begin{figure}[ht]
\begin{center}
\includegraphics[width=6cm]{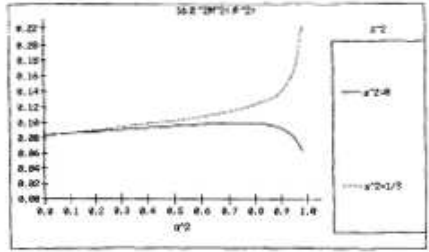}\\
(a)\\
\includegraphics[width=6cm]{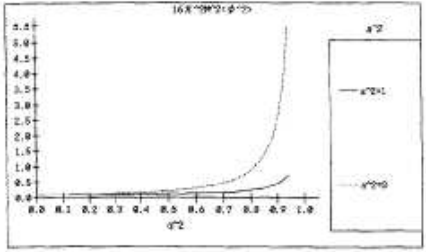}\\
(b)
\caption{(a) $16\pi^2M^2\langle\varphi^2\rangle$ at the horizon is
plotted against
$q^2=Q^2/\{(1+ a^2)M^2\}$ for $a^2=0, 1/3$; (b)
$16\pi^2M^2\langle\varphi^2\rangle$ at the horizon is plottad against
$q^2$ for $a^2=1, 3$.}
\label{f1}\end{center}
\end{figure}

\begin{figure}[ht]
\begin{center}
\includegraphics[width=6cm]{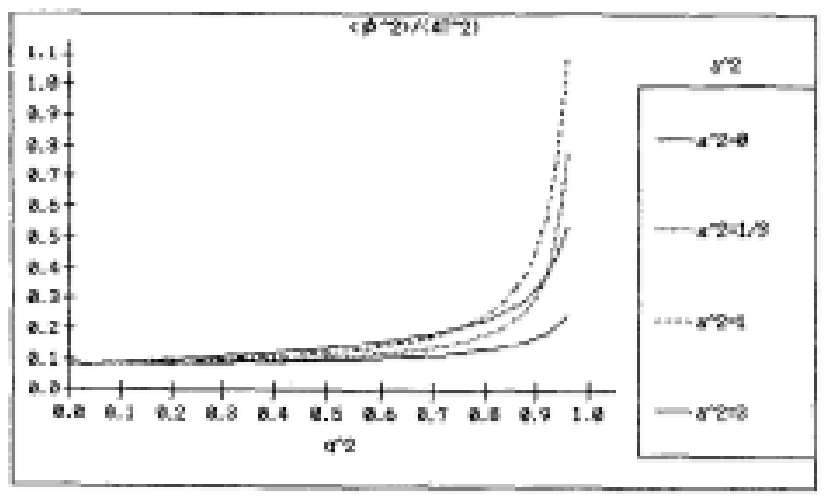}
\caption{$\langle\varphi^2\rangle/(4T_H^2)$ at the horizon is plotted
against $q^2$ for $a^2=0, 1/3, 1, 3$.}
\label{f2}\end{center}
\end{figure}

In the extremal limit, $q\rightarrow 1$, $\langle\varphi^2\rangle$
diverges in general, except for $a=0$ case, i.e., the
Reissner-Nordstr\"om space-time. The ratio
$\langle\varphi^2\rangle/(4T_H^2)$ diverges regardless of the value for
$a$.

We have found that the critical value for $a$ is zero when we pay
attention to the vacuum polarization at the horizon of the maximally
charged dilatonic black hole.

In this paper, we discussed only minimally coupled scalar field. The
couplings to the dilaton field as well as the curvature will change the
behavior of the quantum fields; besides, the connection to the
thermodynamics of black holes and the quantum effects in the other
vacua are to be studied. The relationship to supersymmetry is also of
great interest.


\end{document}